\documentclass{emulateapj}

\slugcomment{}

\shorttitle{AGN feedback at $z\sim2$}
\shortauthors{Cimatti et al.}

\begin{document}


\title{AGN feedback at $z\sim2$ and the mutual evolution of active and
inactive galaxies}

\author{A. Cimatti\altaffilmark{1}, M. Brusa\altaffilmark{1,4}, M. Talia\altaffilmark{1}, M. Mignoli\altaffilmark{2}, G. Rodighiero\altaffilmark{3}, J. Kurk\altaffilmark{4}, P. Cassata\altaffilmark{5}, C. Halliday\altaffilmark{6}, A. Renzini\altaffilmark{7}, E. Daddi\altaffilmark{8}}
\altaffiltext{1}{Dipartimento di Fisica e Astronomia, Universit\`a di Bologna, 
Viale Berti Pichat 6/2, I-30127, Bologna, Italy} 
\email{a.cimatti@unibo.it}
\altaffiltext{2}{INAF, Osservatorio Astronomico di Bologna, Via Ranzani 1, I-40127, Bologna, Italy}
\altaffiltext{3}{Dipartimento di Fisica e Astronomia, Universit\`a di Padova, Vicolo dell'Osservatorio 3, I-35122, Italy}
\altaffiltext{4}{Max-Planck-Institut f\"ur Extraterrestrial Physik, Giessenbachstrasse, 85748 Garching bei M\"unchen, Germany}
\altaffiltext{5}{Aix Marseille Universite, CNRS, Laboratoire d'Astrophysique
de Marseille, UMR 7326, 13388, Marseille, France}
\altaffiltext{6}{23 rue d'Yerres, 91230, Montgeron, France}
\altaffiltext{7}{INAF, Osservatorio Astronomico di Padova, Vicolo dell' Osservatorio 5, I-35122, Italy}
\altaffiltext{8}{Laboratoire AIM, CEA/DSM-CNRS-Universite Paris Diderot, Irfu/Service d'Astrophysique, CEA Saclay, Orme des Merisiers, 91191 Gif-sur-Yvette Cedex, France}
%\altaffiltext{1}{Visiting Astronomer, Cerro Tololo Inter-American Observatory.
%CTIO is operated by AURA, Inc.\ under contract to the National Science
%\altaffiltext{2}{Society of Fellows, Harvard University.}
%\altaffiltext{3}{present address: Center for Astrophysics,
%    60 Garden Street, Cambridge, MA 02138}
%\altaffiltext{4}{Visiting Programmer, Space Telescope Science Institute}
%\altaffiltext{5}{Patron, Alonso's Bar and Grill}

\begin{abstract}
The relationships between galaxies of intermediate stellar mass and 
moderate luminosity active galactic nuclei (AGNs)
 at $1<z<3$ are investigated with the {\it Galaxy Mass Assembly ultra-deep 
Spectroscopic Survey} (GMASS) 
sample complemented with public data in the GOODS-South field. 
Using X-ray data, hidden AGNs are identified in unsuspected star-forming 
galaxies with no apparent signs of non-stellar activity.
In the color--mass plane, two parallel trends emerge during the $\sim$2 Gyr 
between the average redshifts $z\sim2.2$ and $z\sim1.3$: while the 
{\it red sequence} becomes significantly more populated by ellipticals, 
the majority of AGNs with $L(2-10 \ {\rm keV})>10^{42.3}$ erg s$^{-1}$ 
disappear from the {\it blue cloud/green valley} where they were hosted 
predominantly by 
star-forming systems with disk and irregular morphologies. 
These results are even clearer when the rest-frame colors are
corrected for dust reddening.
At $z\sim2.2$, the ultraviolet spectra of active galaxies (including two
Type 1 AGNs) show possible 
gas outflows with velocities up to about -500 km s$^{-1}$ that are not observed 
neither in inactive systems at the same redshift, nor at lower redshifts. 
Such outflows indicate the 
presence of gas that can move faster than the escape velocities of active 
galaxies. These results suggest that feedback from moderately luminous
AGNs (log$L_{X}<44.5$ erg s$^{-1}$) played a key role at $z\gtrsim 2$ 
by contributing to outflows capable 
of ejecting part of the interstellar medium and leading to a rapid decrease 
in the star formation in host galaxies with stellar masses 
$10<$log(${\cal M}/M_{\odot})<$11.
\end{abstract}





\section{Introduction}

The evolution of galaxies is thought to be deeply linked to AGN activity 
because of the tight correlation observed between the masses of present-day
galaxy bulges and their nuclear supermassive black holes (see 
Alexander \& Hickox 2012 for a review). 

AGNs are expected to provide negative feedback processes that can 
suppress star formation on galactic scales through the 
interaction of radiation, winds, and jets with the interstellar 
medium of the host galaxy leading to the ejection and/or heating of a
substantial fraction of the gas (e.g. Di Matteo et al. 2005; Fabian
2012). However, it seems also plausible that the same AGNs can provide 
some positive feedback by triggering or enhancing star formation
(e.g. De Young 1989; Gaibler et al. 2012; Ishibashi, Fabian \& Canning 2013). 

One promising approach is to investigate the relationships between AGN 
activity and galaxy evolution in the critical transformation phase believed 
to occur at $1<z<3$ when a significant fraction of
galaxies moved from the locus of star-forming systems in the 
color -- mass plane (the so called {\it blue cloud}) to the {\it red sequence} 
where spheroidal galaxies with weak or suppressed star formation are located 
mostly at $z \lesssim 1$ (e.g. Cassata et al. 2008; Brusa et al. 2009; 
Silverman et al. 2008; Cardamone et al. 2010; Cameron et al. 2011;
Kocevski et al. 2012). Deep multi-wavelength data are essential to 
address these questions from the observational point of view, and recent 
results suggest that AGN and galaxy evolution 
are deeply related to each other (e.g. Daddi et al. 2007; Mullaney et al. 
2012; Bongiorno et al. 2012; Rovilos et al. 2012; Olsen et al. 2013;
Rosario et al. 2013).

The main purpose of the present study is to investigate the supposed
role AGN negative feedback during the critical cosmic epoch at $1<z<3$.
We adopt $H_0=70$ km s$^{-1}$ Mpc$^{-1}$, $\Omega_{\rm m}= 0.3$, 
$\Omega_{\Lambda}=0.7$.

\section{The GMASS sample and X-ray data}

The present study is based on the {\it Galaxy Mass Assembly ultra-deep
Spectroscopic Survey} (GMASS) sample, in which galaxies were 
selected from a region of the GOODS-South/Hubble Ultra Deep Field region 
with the only criterion to have {\it Spitzer} IRAC $4.5 \mu$m magnitudes 
brighter than 23.0 (AB) (Kurk et al. 2013; K13 hereafter). As shown
in K13, the selection at 
4.5$\mu$m ensures sensitivity to stellar mass, a favorable $k$-correction 
up to z$\sim$3, less influence by dust extinction, and represents a
major difference compared to previous works in this sky region where
studies on AGNs were mostly based on X-ray selected samples (see
previous references). A fraction of 
galaxies have very deep optical spectroscopy with integration times up to
$>$30 hours (K13). Photometric spectral energy distributions (SEDs) from 
\emph{U}- to the IRAC $8\mu m$-band were used to estimate photometric 
redshifts (photo-$z$) for galaxies without spectroscopic redshifts and 
to derive galaxy properties adopting the stellar population models of 
Maraston (2005), the Kroupa (2001) initial mass function (IMF), and solar 
metallicity (K13). These SEDs were also used to derive the rest-frame 
(U-B) colors (standard Johnson filters) with the method described 
by Ilbert et al. (2005). 

For the present study, 266 galaxies were selected from the GMASS sample
applying two selection criteria: $1<z<3$ and a cut in stellar mass, 
log(${\cal M}/M_{\odot})>$10, in order to ensure a high 
completeness in mass across the whole redshift range (see Cassata et al. 
2008). At $1<z<3$, 55\% of the sample have reliable spectroscopic redshifts 
when GMASS spectroscopy is complemented with the public ESO VLT GOODS-South 
spectra (Balestra et al. 2010). For the
remaining 45\% of the sample, the photo-$z$ described in K13 were used. 

Public {\it Hubble Space Telescope} (HST) near-infrared imaging (WFC3
+ F160W filter) taken in the framework of the ERS (Windhorst et al. 2010) 
and CANDELS (Koekemoer et al. 2011; Guo et al. 2013) surveys was exploited 
to assign one of the following visual (rest-frame optical) morphological 
classes to each galaxy:  elliptical, compact, disk, irregular, and faint 
(Fig. 1 and Table 1; see also Talia et al. 2013).
The galaxies classified as compact are small round-like objects, whereas 
the faint galaxy class consists of objects too weak (or invisible) for a 
reliable classification. 

\begin{figure}
\centering
\includegraphics[clip=true, width=20mm]{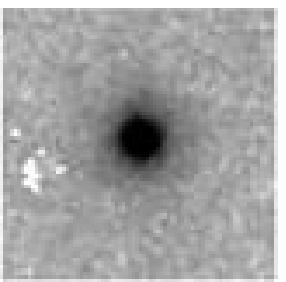}
\includegraphics[clip=true, width=20mm]{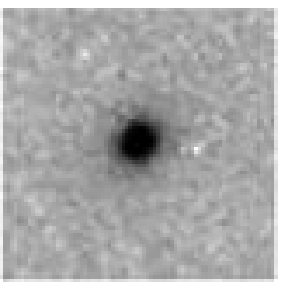}
\includegraphics[clip=true, width=20mm]{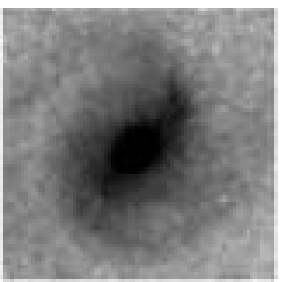}
\includegraphics[clip=true, width=20mm]{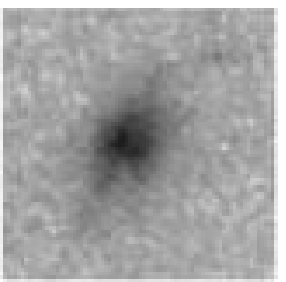}
\includegraphics[clip=true, width=20mm]{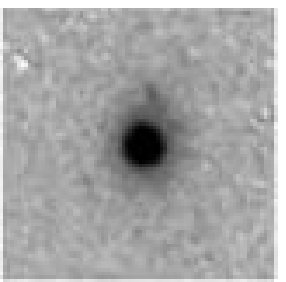}
\includegraphics[clip=true, width=20mm]{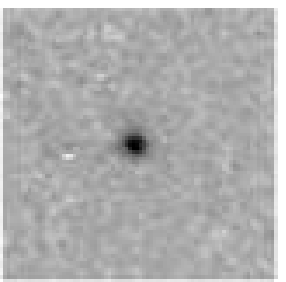}
\includegraphics[clip=true, width=20mm]{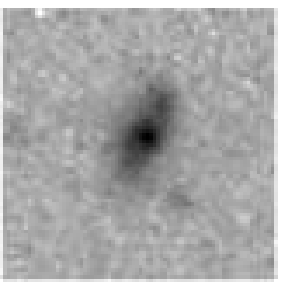}
\includegraphics[clip=true, width=20mm]{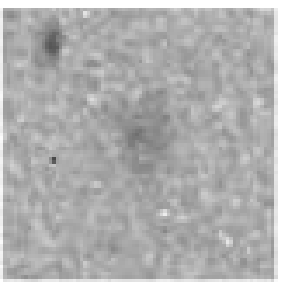}
\caption{Examples of morphological classes (HST, WFC3+F160W imaging,
$4^{\prime\prime}\times4^{\prime\prime}$). From left to right: elliptical 
($z=1.31$), 
elliptical ($z=1.19$), disk ($z=1.09$), irregular ($z=1.03$), 
elliptical ($z=1.91$), compact ($z=2.51$), disk ($z=1.84$), faint
($z=2.45$).} 
\end{figure}							 

Our selected sample of 266 galaxies was cross-correlated with the catalog 
of 740 X-ray sources of Xue et al. (2011, X11 hereafter) based on the 
{\it Chandra} 4 Msec public data of the Chandra Deep Field South 
(CDF-S), returning 58 matches within a search radius of 1$^{\prime\prime}$.
The rest frame 2-10 keV luminosities ($L_X$ hereafter)
were derived from the absorption-corrected rest-frame 0.5-8 keV luminosities
tabulated in X11 and assuming a $\Gamma$=1.8 power-law.
Neutral hydrogen column densities (N$_{\rm H}$) were retrieved from 
Bauer et al. (private communication). The 4Ms CDFS exposure is complete
down to log$L$(0.5-8 keV)$\sim 42.5$ erg s$^{-1}$ across the entire 
redshift range explored in this work (see Figure 16b in X11), 
which translates into log$L_X \sim 42.3$ erg s$^{-1}$ 
for Compton-thin AGNs (N$_{\rm H}$$<10^{23}$ cm$^{-2}$). This is also the
threshold above which AGN activity dominates over star formation 
(e.g. Ranalli et al. 2003; Bauer et al. 2004). 

The Chandra data were also used to derive average X-ray properties of
galaxies (such as fluxes and average obscuration) 
by using a stacking technique based on CSTACK\footnote{http://cstack.ucsd.edu/cstack/}(v3.0) 
recently modified to include the full 4Ms exposure. In order to minimize 
any contamination of detected sources in the stacked signal, all objects 
with a known X-ray source within $10^{\prime\prime}$ of the centroid were 
excluded. Furthermore, the stacking analysis has been made within 
8$^{\prime}$ of the CDFS field center in order to maximize the Chandra PSF.

\subsection{General properties of the sample}

Our selected sample probes the regimes of intermediate galaxy 
stellar masses ($10<$log(${\cal M}/M_{\odot})<$11) and moderate luminosity 
AGNs (log$L_{X}<44.5$ erg s$^{-1}$) (Fig. 2).

\begin{figure}
\begin{center}
\includegraphics[width=0.98\linewidth]{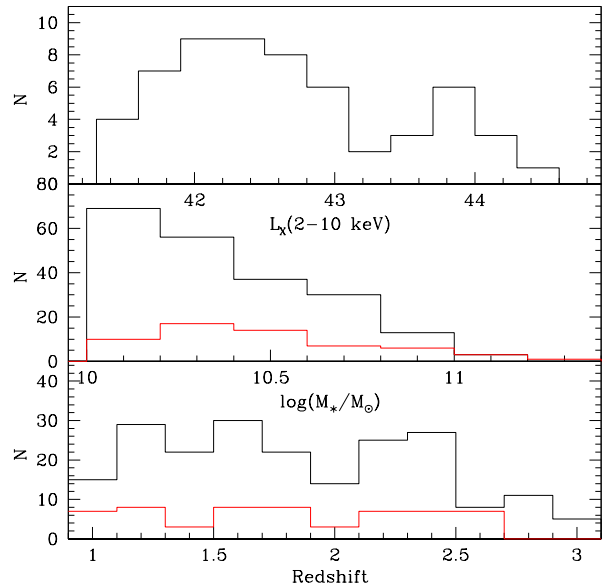}
\caption{The distributions or redshifts (spectroscopic-$z$ + photo-$z$),
stellar masses and X-ray luminosities. Black and red: galaxies without 
and with X-ray emission respectively.}
\label{}
\end{center}
\end{figure}

The fractions of X-ray sources at $1<z<3$ are 22\% and 13\%  
down to log$L_{X}=41.3$ and log$L_{X}>42.3$, respectively, 
which are consistent with previous works (e.g. Xue et al. 2010; Olsen 
et al. 2013)
when the increasing fraction of AGNs with stellar mass is taken into account. 
If the sample is divided into two redshift bins nearly equally populated 
($1<z<1.7$, N=122, $z_{med}=1.31$, $1.7<z<3$, N=144, $z_{med}=2.25$), 
the fraction of X-ray sources with log$L_{X}>42.3$ depends 
on redshift, increasing from 7\% at $1<z<1.7$ to 18\% at $1.7<z<3$ (see
also Brusa et al. 2009). 

\begin{table}
\caption{Statistics of morphological classes}
\begin{tabular}{l l l l l l l}
\tableline
 & Elliptical & Compact & Disk & Irregular & Faint\\
 & & & & & & \\
$1<z<1.7$ & & & & & & \\
All&40\% (49)&0\% (0) &43\% (53) &16\% (20) &0\% (0) \\
No X-rays & 37\% (36) & 0\% (0) & 46\% (44) & 17\% (16) & 0\% (0)\\
$L_X>10^{41.2}$ & 50\% (13) & 0\% (0) & 35\% (9) & 15\% (4) & 0\% (0)\\
$L_X>10^{42.3}$ & 62\% (5) & 0\% (0) & 25\% (2) & 12\% (1) & 0\% (0)\\ \\
$1.7<z<3$ & & & & & & \\
All&17\% (25)&10\% (15) &17\% (24) &47\% (68) &8\% (12) \\
No X-rays & 14\% (16) & 10\% (11) & 16\% (18) & 51\% (57) & 9\% (10) \\
$L_X>10^{41.2}$ & 28\% (9) & 12\% (4) & 19\% (6) & 34\% (11) & 6\% (2) \\
$L_X>10^{42.3}$ & 23\% (6) & 15\% (4) & 19\% (5) & 35\% (9) & 8\% (2) \\
\tableline
\end{tabular}
\tablenotetext{a}{The fractions (\%) and number (in brackets) of morphological 
classes.}
\end{table}

\section{Mutual galaxy--AGN evolution}

An intriguing result emerges when the evolutions of galaxies and AGNs 
with log$L_{X}>42.3$ erg s$^{-1}$ are compared in the rest-frame 
color -- stellar mass plane at $1.7<z<3$ and $1<z<1.7$ (Fig. 3).

Using the (U-B) colors uncorrected for dust reddening, at 
$1.7<z<3$, the majority of galaxies 
($\sim$78\%) do not lie on the {\it red sequence}. AGNs are mostly
located in the {\it blue cloud} and {\it green valley} regions (92\%),
and hosted predominantly by star-forming galaxies with irregular and disk 
morphologies (see also Rosario et al. 2013). A smaller fraction of AGNs is 
hosted by ellipticals and compact galaxies (Table 1).

\begin{figure}
\begin{center}
\includegraphics[width=0.98\linewidth]{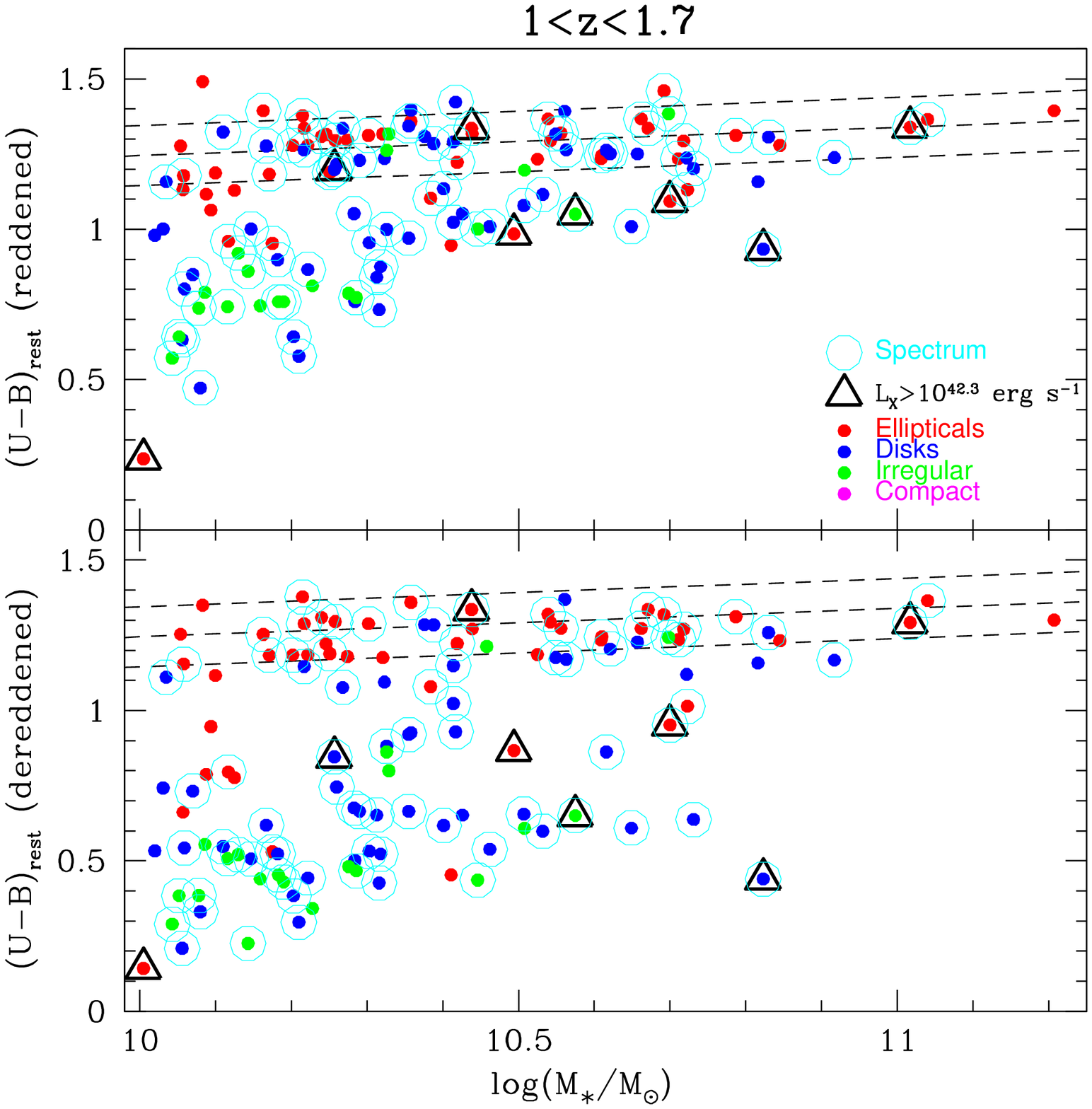}
\includegraphics[width=0.98\linewidth]{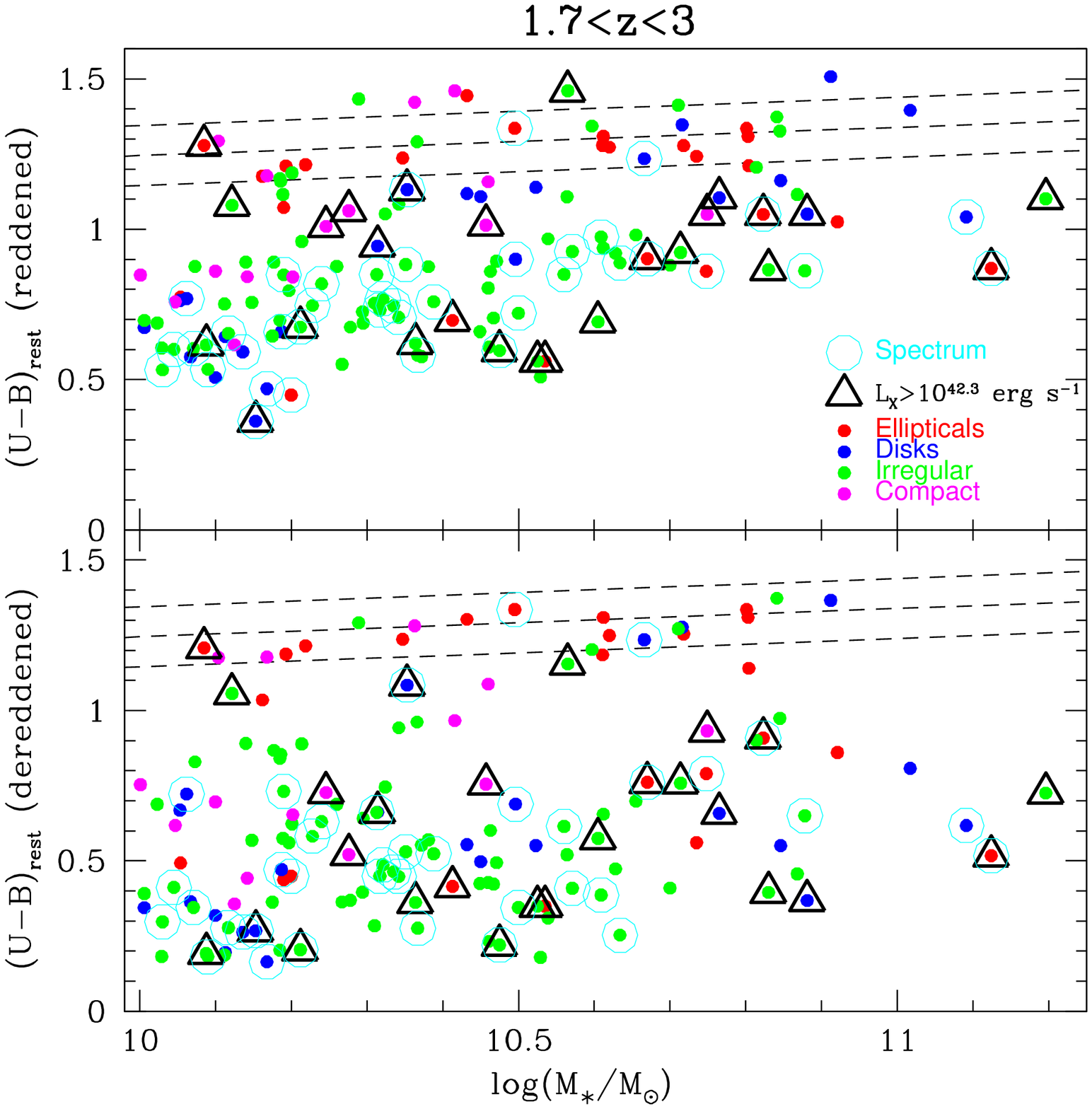}
\caption{The rest-frame color -- stellar mass diagrams at $1<z<1.7$ (top panel) 
and $1.7<z<3$ (bottom panel). In each panel, both dust-reddened and 
dereddened colors are shown. Dashed lines: the {\it 
red sequence} defined by Cassata et al. (2008). Cyan open circles:
optical spectrum and spectroscopic-$z$ available. Triangles: X-ray
sources with $L_X>10^{42.3}$ erg s$^{-1}$.}
\label{}
\end{center}
\end{figure}

At $1<z<1.7$ the {\it red sequence} becomes more established than
in the higher redshift bin, containing $\sim$56\% of the whole galaxy 
population (compared to $\sim$22\% at $1.7<z<3$). 
The fraction of morphological ellipticals increases to 
$\sim$40\% (compared to $\sim$17\% at $1.7<z<3$), the majority
of which ($\sim$77\%) are located on the {\it red sequence} (vs. 
$\sim$44\% at $1.7<z<3$). The ellipticals belonging to the 
overdensity at $z=1.61$ (Kurk et al. 2009) represent only 16\% of the 
total number of ellipticals in the {\it red sequence}. The {\it blue 
cloud} becomes mostly populated by disk and irregular galaxies, while 
no compact and faint objects are observed. Compared to $1.7<z<3$,
the AGNs become rarer and are hosted mainly by elliptical and disk galaxies. 
The fraction of AGNs in the {\it blue cloud} decreases from $\sim$21\% 
at $1.7<z<3$ to $\sim$9\% at $1<z<1.7$.

The above results are clearer when the (U-B) colors are
dereddened using the dust extinction (A$_{\rm V}$) derived from 
the photometric SED fitting (K13) and adopting the Calzetti (2001)
extinction curve. Fig. 3 shows that
the {\it red sequence} can be more clearly distinguished from the 
{\it blue cloud} (see also Cardamone et al. 2010), 
and populated by a fraction of morphological ellipticals increasing from
$\sim$52\% at $1.7<z<3$ to $\sim$81\% at $1<z<1.7$.

These evolutionary patterns suggest a scenario where the evolutions
of galaxies and AGNs are closely related at $z \gtrsim 2$, when
AGNs were hosted by star-forming systems, after which  
($\sim$2 Gyr later) the number of AGNs decreased and the 
number of {\it red sequence} galaxies increased.
In this picture, the compact galaxies hosting an AGN
may represent a transitional phase between the concomitant
star-forming and AGN activities in the {\it blue cloud}, and the
subsequent {\it quenching} of star formation and morphological 
transformation, thus populating the {\it red sequence} while the 
fraction of AGN hosts drops rapidly.

\section{Clues on hidden AGNs at $1.7<z<3$}

The properties of active and inactive galaxies were investigated 
using their stacked optical spectra. At $1<z<1.7$ and $1.7<z<3$, the fractions
of galaxies with available spectra are 95/122 and 44/144 respectively,
with a spectroscopic completeness for the X-ray galaxies of
85\% and 34\%. The stacking of optical spectra was done in three steps: 
(i) the individual spectra were smoothed at the lowest resolution (i.e. 
R$\sim$200 of the VIMOS LR grism), (ii) each spectrum was rescaled by the 
average flux in a common spectral region, (iii) the spectra were averaged
without any weights. In the AGN sample, we removed the 3 objects 
classified as Type 1 AGN (ID2043, ID1155 and ID1350; Fig. 4). 

At $1.7<z<3$, the stacked spectrum of active (i.e. detected individually
in X-rays) and inactive galaxies was made with 9 ($z_{med}$=2.0) and 30 
($z_{med}$=2.25) spectra respectively. 
The spectrum of the active galaxy ID 2171 ($z$=2.145) was excluded 
from the stacking because of its strong emission lines, which differ
markedly from the other galaxies (Fig. 4). 

Compared to inactive galaxies, the stacked spectrum of active galaxies 
shows no Ly$\alpha$ emission and stronger C III]$\lambda$1909 emission
(Fig. 4). Moreover, some interstellar medium (ISM) absorption lines (e.g. 
Si II$\lambda$1260, O I+Si II$\lambda$1303, Al II$\lambda$1670) have 
equivalent widths (EWs) systematically larger than in inactive galaxies by 
a factor of $\sim$2 (EW$_{\rm rest}\sim$2-4~\AA). 

The X-ray data were then used to assess whether the AGN origin of the X-ray 
emission (assumed in Section 2 based simply on the criterion of
log$L_{X}>42.3$ erg s$^{-1}$)
is indeed correct. 

The 9 galaxies at $1.7<z<3$ 
with spectroscopic redshifts that are detected individually 
in X-rays have an average luminosity of 4$\times10^{42}$ erg s$^{-1}$ and 
a hardness ratio HR$\sim$0 (HR=H-S/H+S, with H and S being the count 
rates in the hard and soft bands respectively), that is too flat to be 
explained by star formation only and consistent with an absorbed power-law 
with $\Gamma=1.8$, which is typical of an obscured AGN. These results remain 
unchanged when only the 6 galaxies with log$L_{X}>42.3$ erg s$^{-1}$ 
are used, or the stacking includes also the X-ray galaxies with 
photo-$z$ and all $L_X$ (for a total of 24/26 objects used by CSTACK). 
The individual galaxy ID 2171 (not included in the stacking of optical 
spectra discussed later) has X-ray properties typical of a Compton thick 
AGN with $L_X= 1\times10^{43}$ erg s$^{-1}$, HR$>$0.85 and N$_{\rm H}$$\geq 2\times
10^{24}$ cm$^{-2}$.

On the basis of SED fitting (K13), UV spectral slopes (Talia et al. 2012) and 
infrared luminosities (Talia et al., in preparation), the 
median star formation rate (SFR) of the X-ray emitting galaxies is consistently 
$\sim$100 M$_\odot$ yr$^{-1}$. According to Ranalli et al. (2003), this SFR 
corresponds to $L_X \sim 5 \times10^{41}$ erg s$^{-1}$, i.e. $\sim$1 dex less 
than the observed total X-ray luminosity, indicating a negligible 
contribution of star formation to the observed $L_X$. 

In stark contrast, the X-ray stacked flux of the 30 inactive galaxies (i.e.
individually undetected in X-rays) with spectroscopic redshifts is significant only 
in the soft-band (0.5-2 keV). 
Assuming an unobscured spectrum with $\Gamma=2$, 
the detected soft-band flux implies an average  
$L_X \sim10^{41.1}$ erg/s at mean redshift $z=2.25$ and can be 
ascribed to SFR$\sim25$ M$_\odot$yr$^{-1}$. 
These results remain unchanged when the X-ray stacking 
includes also the galaxies with photo-$z$ (87/112 used by CSTACK). 

It is remarkable that, despite the clear presence of AGNs (as
shown by the X-ray properties), the UV
spectra of active galaxies at $1.7<z<3$ do not show (except for ID 2171) any clear features 
of non-stellar activity,
particularly when compared to inactive galaxies of similar redshift and 
stellar mass. One possibility is that the AGNs are hosted by more obscured 
galaxies with denser ISM, which would explain the larger EWs of some ISM 
absorption lines and the lack of Ly$\alpha$ emission.

\begin{figure}
\begin{center}
\includegraphics[width=0.98\linewidth]{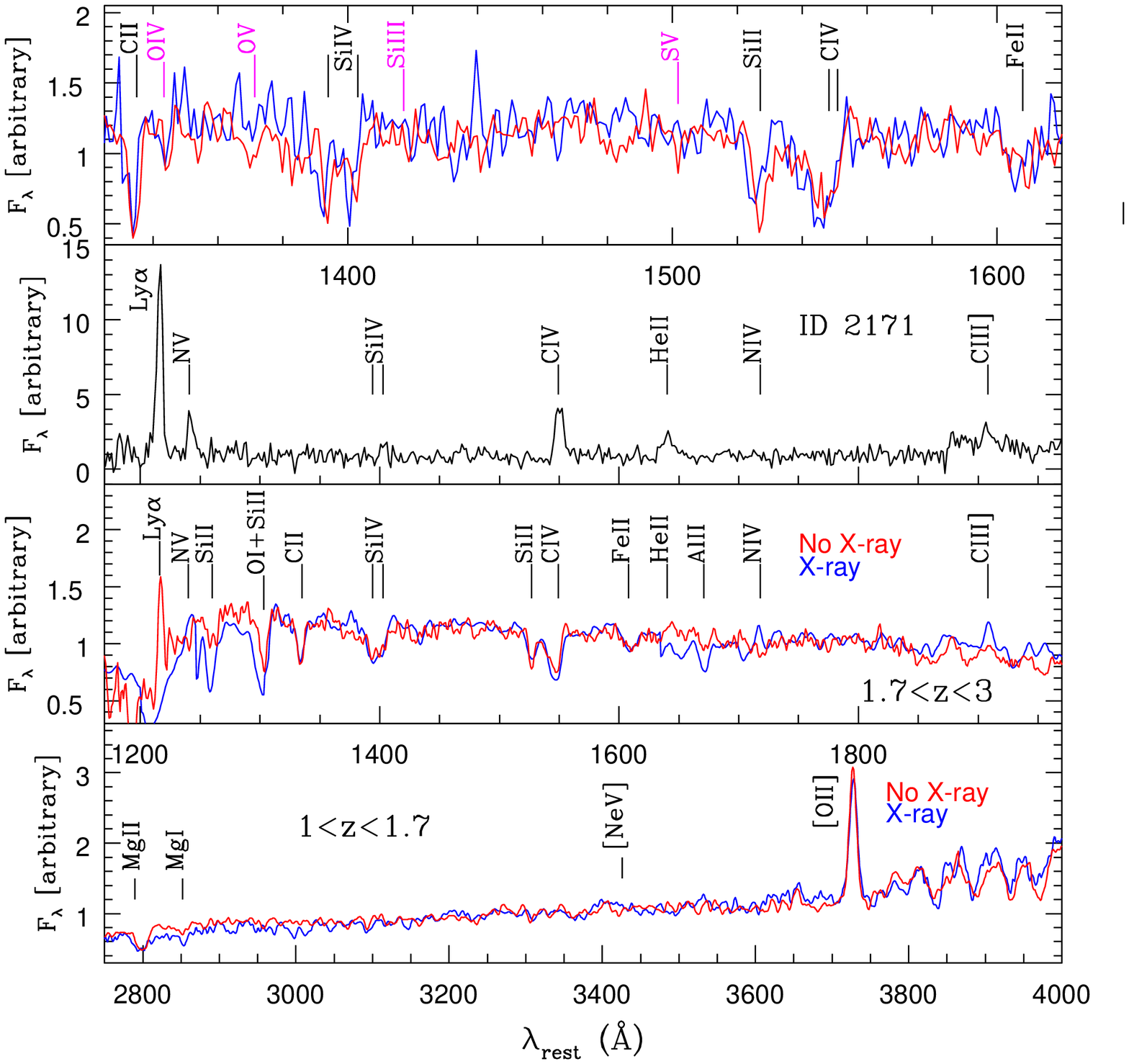}
\includegraphics[width=0.98\linewidth]{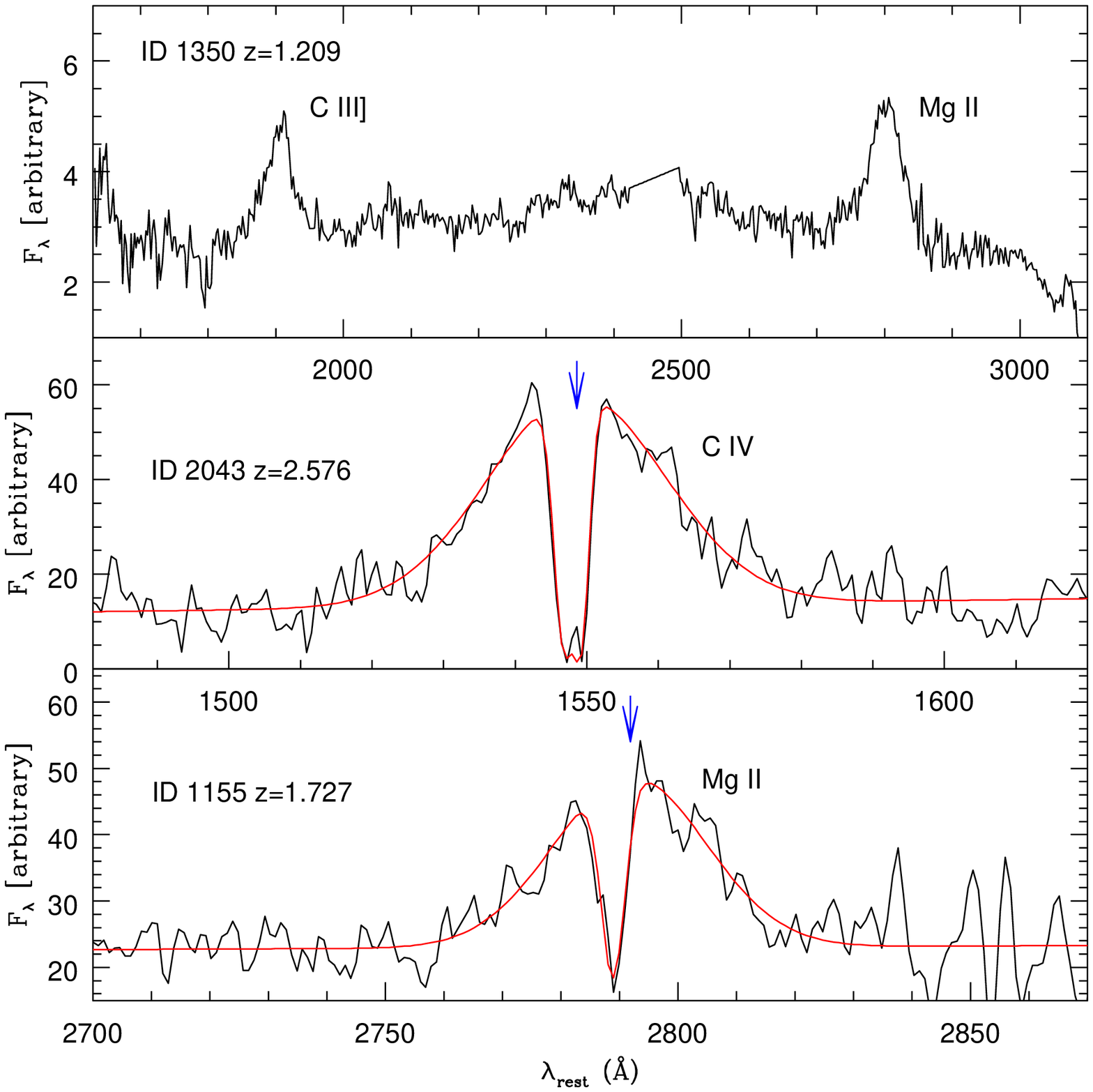}
\caption{{\it Top panel}. From bottom to top: the stacked spectra
at $1<z<1.7$, at $1.7<z<3$, the spectrum of ID 2171, and
the stacked GMASS-only spectra (R$\sim$600) at $1.7<z<3$
placed at the systemic reference defined by photospheric lines.
Red and blue spectra are relative to inactive 
and active galaxies respectively.
{\it Bottom panel}: the spectra of Type 1 AGNs and their best fits
(red). Blue arrows: the wavelength of the emission line centroid.}
\label{}
\end{center}
\end{figure}

\section{Searching for outflows at $1.7<z<3$}

We searched for ISM gas ouflows in the stacked spectra at $1.7<z<3$ 
in order to further investigate the differences between active and 
inactive galaxies.

For this purpose, only the subset of galaxies with the highest 
resolution spectra from GMASS (R$\sim$600; K13) were used in order
to maximize the radial velocity accuracy and to detect the weak 
photospheric lines (C III$\lambda$1247, O IV$\lambda$1343, O V$\lambda$1371, 
Si III$\lambda$1417, S V$\lambda$1501; see Leitherer et al. 2011) 
whose median wavelength was used in each stacked spectrum to define the 
systemic (zero velocity) reference (topmost spectra in Fig. 4).  

In the stacked spectrum of the 24 inactive galaxies, 
the main ISM absorption lines (Si II$\lambda$1260, O I$\lambda$1302, 
C II$\lambda$1334, Si II$\lambda$1526, Si IV$\lambda\lambda$1393,1402, 
Fe II$\lambda$1608, Al II$\lambda$1670) show a median blueshift of -70$\pm$100
km s$^{-1}$, consistently with other results on inactive galaxies
(e.g. Shapley et al. 2003; Steidel et al. 2011; Talia et al. 2012). 

In contrast, for the stacked spectrum of 7 active galaxies, we detect
larger blueshifts (with $\Delta V$ between -600 and -120 km s$^{-1}$) for 
the same lines, 
with a median value of -340$\pm$150 km s$^{-1}$. 
Similar blueshifts (ascribed to outflows) have 
been observed also in other active galaxies at $1<z<3$ (Nesvadba et al. 2006;
Hainline et al. 2011; Harrison et al. 2012). 

For the typical size ($r_e\sim 5$ kpc) of the active galaxies of Fig.4,
the expected escape velocity is $V_{\rm esc}=\sqrt{2GM/r} \sim$ 350 km
s$^{-1}$, where $log(M/M_{\odot})=10.9$ is the sum of the median stellar 
mass (log(${\cal M}/M_{\odot})=10.4$) and the total gas mass (H I+H$_2$) 
(log($M_{gas}/M_{\odot})\sim$10.7; Magdis et al. 2012, eq. 27)
This $V_{\rm esc}$ is comparable to the measured median 
velocity, and lower than the highest observed speeds. The real
$\Delta V$ could be even larger because the blueshifts account only 
for the radial velocity component. 

Although the presence of OB star photospheric lines implies a role of 
stellar feedback, the larger velocities observed in active galaxies 
suggest an additional contribution from AGNs to the energy of gas outflows. 
Deeper integral field spectroscopy would be useful to confirm these
results and study the geometry of the outflows, as e.g., in Newman et
al. (2012).

We also searched for outflows in the spectra of the two Type 1 AGNs 
identified at 
$1.7<z<3$ (ID 1155 at $z=1.727$ and ID 2043 at $z=2.576$; K13). Both objects 
have log$L_X \sim$44 erg s$^{-1}$, heavy to moderate obscuration
(N$_{\rm H} =1.3\times10^{23}$ 
and $1\times10^{21}$ cm$^{-2}$), and display broad emission lines 
(FWHM$\sim$3200 km s$^{-1}$ for Mg II$\lambda$2800, FWHM$\sim$5600 km s$^{-1}$ 
for C IV$\lambda$1549). Narrow (FWHM$\sim$500 km s$^{-1}$) absorption 
lines associated with MgII$\lambda$2800 and CIV$\lambda$1549 are present
in ID 1155 and 2043 respectively (Fig. 4). A three-component best fit 
of the continuum, emission, and absorption lines indicates that the
absorptions are blueshifted by $-311 \pm 50$ km s$^{-1}$ (MgII)
and $-140 \pm 50$ km s$^{-1}$ (CIV) relative to the peak of the emission 
lines (adopted as the systemic reference) (Fig.4).
These results suggest that gas outflows are also present in Type 1 AGNs
at $1.7<z<3$.

\section{The fading activity at $1<z<1.7$}

Moving to lower redshifts, the 20 galaxies at $1<z<1.7$ with spectroscopic 
redshifts and individual X-ray detection have an average HR$\sim$0 compatible  
with an absorbed power-law with $\Gamma=1.8$ and column density 
N$_{\rm H}\sim3\times 10^{22}$ cm$^{-2}$, typical of an obscured AGN.
These results do not change when either the 6 galaxies 
with log$L_{X} >42.3$ erg s$^{-1}$ are used, or when 
galaxies with photometric redshifts are included (24/26 used by CSTACK). 

However, despite the clear presence of AGNs, the median X-ray luminosity 
(above the completeness limit of log$L_X>42.3$ erg s$^{-1}$) of active 
galaxies at $1<z<1.7$ (log$L_X$=42.85 erg s$^{-1}$) becomes lower by a 
factor of $\sim$2 than at $1.7<z<3$ (log$L_X$=43.18 erg s$^{-1}$), which
implies, together with the smaller AGN fraction (Section 3), that
the AGN energy release declines with redshift.

In contrast, for inactive galaxies at $1<z<1.7$ (i.e. without 
individual X-ray detection) with spectroscopic redshifts, a significant 
signal was detected with CSTACK (48/73 objects) only in the soft-band.
The detected soft-band flux of $\sim 1.2\times10^{-17}$ cgs (assuming an 
unobscured spectrum with $\Gamma=2$) implies $L_X \sim10^{40.8}$ erg/s 
for the average redshift ($z=1.32$) of these galaxies.  

If this luminosity is solely due to star formation, according to Ranalli 
et al. (2003), it implies an average SFR$\sim$14 M$_\odot$yr$^{-1}$, 
broadly consistent with the average(median) SFR$\sim$30(5) 
M$_\odot$yr$^{-1}$ derived for these galaxies from the photometric SED 
fitting (K13). The non-detection of hard X-rays also indicates a negligible 
contamination of obscured and low-luminosity AGNs in our sample of intermediate 
mass galaxies (but see Olsen et al. 2013 for a higher stellar mass 
sample). These results do not 
change significantly when the X-ray stacking is performed including 
galaxies with photometric redshifts (64/96 used by CSTACK). 

At $1<z<1.7$, the optical spectra of active (N=20, $z_{med}$=1.22, all
$L_X$) and inactive (N=73, $z_{med}$=1.29) galaxies have very 
similar properties (Fig. 4). The absence of [Ne V]$\lambda$3426 emission 
(expected in case of AGN photoionization) in the stacked spectrum of 
X-ray galaxies may be due to attenuation by dust extinction and/or the
gas density being higher than the critical density of this line 
(see Mignoli et al. 2013). 

Even with the high resolution of the GMASS spectra, the search for gas 
outflows in this redshift range is uncertain because it is 
impossible to clearly distinguish either the [O II]$\lambda$3727 and 
Mg II$\lambda$2800 doublets, or the 
ISM and photospheric absorption components of Mg II.
For instance, adopting 3727.5~\AA~ as the systemic centroid of the unresolved 
[O II]$\lambda$3726+3729 doublet (as in Bradshaw et al. 2013), the stacked 
spectra of active and inactive galaxies show Mg II$\lambda$2800 velocity 
offsets in the range of -200 to -50 km s$^{-1}$ with a typical uncertainty 
of $\sim$100 km s$^{-1}$.

In contrast to those identified at $1.7<z<3$, the single Type 1 AGN
(ID 1350, $z=1.209$, log$L_X$=43.97 erg s$^{-1}$; Fig. 4) appears
to have no intrinsic absorption and properties that are indeed typical of 
an unabsorbed Type I AGN with a blue continuum. 

 Interpreted with our findings in Section 3, these results
suggest that the 
AGN activity and its effects on the host galaxies have faded gradually
from $1.7<z<3$ to $1<z<1.7$.

\section{Summary}

We have presented evidence of a link between the migration of galaxies from 
redshift $z\sim2$ to $z\sim1$ onto the red sequence and a parallel decrease 
in the activity of AGNs with $L_X>10^{42.3}$ erg s$^{-1}$. At $z\sim2$, 
the AGNs often remain hidden in unsuspected star-forming galaxies. 

We tentatively detect gas outflows at speeds of up to about -500 km s$^{-1}$, 
which are comparable to or larger than the galaxy escape velocities, 
and present only in active galaxies at $z\sim2$. This suggests that AGN 
feedback, in addition to star formation, contributes to these gas outflows, 
removing some fraction of the gas permanently from the galaxy, 
leading to so-called star-formation {\it quenching}, and allowing
a fraction of galaxies to migrate onto the {\it red sequence}.

Deeper spectroscopy is needed to confirm 
and extend these results to larger samples.
The synergy of future massive imaging-spectroscopy surveys (e.g. Euclid; 
Laureijs et al. 2011) and X-ray missions such as {\it eROSITA} (Merloni
et al. 2012) and {\it Athena+} (Georgakakis et al. 2013) will be crucial 
to fully unveil the evolutionary links between galaxies and AGNs.

\acknowledgments

We thank the anonymous referee for the constructive comments, Takamitsu 
Miyaji for helping with CSTACK, Franz Bauer for the information on 
N$_{\rm H}$, and Luca Ciotti, Natascha F\"orster-Schreiber and Simon
Lilly for useful discussions. AC and MM acknowledge the support from grants 
ASI n.I/023/12/0 and MIUR PRIN 2010-2011, and MB from the FP7 Career 
Integration Grant ``eEASy'' (CIG 321913).


%% in parentheses, after the keyword, but they will not be verified.

{\it Facilities:} \facility{ESO VLT (FORS2,VIMOS)}, \facility{HST (ACS,WFC3)}, 
\facility{CXO (ACIS)} \facility{Spitzer (IRAC)}.




\clearpage

%% the file, uncomment the \plotone, \plottwo, and
%% If you need a layout that cannot be achieved with \plotone or
%% \plottwo, you can invoke the graphicx package directly with the
%% \includegraphics command or use \plotfiddle. For more information,
%% please see the tutorial on "Using Electronic Art with AASTeX" in the
%% documentation section at the AASTeX Web site, http://aastex.aas.org/
%%
%% The examples below also include sample markup for submission of
%% supplemental electronic materials. As always, be sure to check
%% the instructions to authors for the journal you are submitting to
%% for specific submissions guidelines as they vary from

%% This example uses \plotone to include an EPS file scaled to
%% 80% of its natural size with \epsscale. Its caption
%% has been written to indicate that additional figure parts will be
%% available in the electronic journal.

\end{document}